\documentclass[
preprint,
superscriptaddress,
amsmath,
amssymb,
aps,prx]{revtex4-2}

\usepackage{graphicx}
\usepackage{dcolumn}
\usepackage{bm}
\usepackage{xcolor}
\usepackage[mathlines]{lineno}
\usepackage[normalem]{ulem}
\usepackage[
		colorlinks = true,
            linkcolor = blue,
            urlcolor  = blue,
            citecolor = blue,
            anchorcolor = blue
            ]{hyperref}         
\newcommand\rxout{\bgroup\markoverwith{\textcolor{red}{\rule[.5ex]{2pt}{.6pt}}}\ULon}

\usepackage{adjustbox}  
\begin{document} 

\title{Seven-octave ultrabroadband metamaterial absorbers via Q-weighted mode density modulation}



\author{Nengyin Wang}
\thanks{These authors contributed equally to this work.}
\affiliation{Institute of Acoustics, Tongji University, Shanghai 200092, China}

\author{Sibo Huang}
\thanks{These authors contributed equally to this work.}
\affiliation{Department of Electrical Engineering, City University of Hong Kong, Hong Kong, China}

\author{Zhiling Zhou}
\thanks{These authors contributed equally to this work.}
\affiliation{Institute of Acoustics, Tongji University, Shanghai 200092, China}

\author{Din Ping Tsai}
\email{dptsai@cityu.edu.hk}
\affiliation{Department of Electrical Engineering, City University of Hong Kong, Hong Kong, China}

\author{Jie Zhu}
\email{jiezhu@tongji.edu.cn}
\affiliation{Institute of Acoustics, Tongji University, Shanghai 200092, China}

\author{Yong Li}
\email{yongli@tongji.edu.cn}
\affiliation{Institute of Acoustics, Tongji University, Shanghai 200092, China}

\date{\today}


\begin{abstract}
Absorption is a crucial parameter in shaping wave propagation dynamics, yet achieving ultra-broadband absorption remains highly challenging, particularly in balancing low-frequency and broad bandwidth. Here, we present a metamaterial absorber (MMA) capable of achieving simultaneous spectral coverage across a seven-octave range of near-perfect absorption from 100 Hz to 12,800 Hz by engineering the quality-factor-weighted (Q-weighted) mode density. The Q-weighted mode density considers mode density, resonant frequencies, radiative loss, and intrinsic loss of multiple resonant modes, providing a comprehensive approach to govern broadband absorption properties. By optimizing the number of resonant modes and managing intrinsic losses, our approach achieves an intensive Q-weighted mode density across an ultra-wide bandwidth, enabling ultra-broadband absorption with high efficiency. These findings significantly advance the bandwidth capabilities of state-of-the-art MMAs and pave the way for the development of ultra-broadband metamaterial devices across various wave systems.
\end{abstract}

\maketitle 
 
\newpage
\section{Introduction}
Absorption is an essential physical phenomenon where wave energy is transformed into other types within a medium or its interfaces. It plays a crucial role in determining the propagation dynamics of sound waves, electromagnetic waves, and beyond. For instance, by judiciously modulating the absorption in non-Hermitian systems, the accompanying absorption can support effective wave manipulation and induce exotic phenomena such as coherent perfect absorption \cite{PK2019Nature, WC2021Science}, asymmetric transmission \cite{SQ2021Science, WangX2019PRL} and reflectionless scattering modes \cite{JX2023NP, SJ2023SciAdv}.  

High-efficiency broadband absorption, in particular, has been a long-standing pursuit due to its significant scientific value and engineering potential
\cite{ZhouZL2022NSR, Qusc2022SA, QuS2021PNAS, WuZ2022AM, XiaY2022AFM, HuangSB2023PRApplied, HuangY2018Nanoscale, WangBX2023AFM, WangNY2023IJEM,YangM2017MH}. Conventional non-resonant resistive materials exhibit fine performance at the cost of bulky volume, while the absorption properties are lack of tunability \cite{YangM2017ARMR, HuangSB2023PRApplied, WangBX2023AFM}. Recently, metamaterial absorbers  (MMAs) have demonstrated exceptional wave manipulation capabilities for customized absorption characteristics by harnessing the resonances of subwavelength structures, which provide more degrees of tuning freedom \cite{CSA2016NRM,Zheludev2012NM, AssouarB2018NRM, LiangL2022AM, ZhouY2023NanoL}. However, the dispersion nature of resonances strongly restricts the operating bandwidth \cite{BliokhK2008RMP, HuangLJ2023NRP, HelmholtzHuang, WangX2019PRL}, confining the highly efficient absorption to narrow-band ranges \cite{LNI2008PRL,LiuX2010PRL, ShenX2012APL, Ma2014NM, AVS2015PRX,LiY2016APL, WangD2023Nanoscale, HeL202B4MRB}. 

To make breakthroughs in the bandwidth, various physical mechanisms have been investigated. The most common design strategy is based on the mode density by coupling local resonances as many as possible. 
In this case, a wide operating bandwidth heavily requires a substantial number of resonant modes \cite{YangM2017MH,ZhouZL2022NSR,HuangSB2023PRApplied, QuS2022PRApplied,FRH2020AM,GW2016OE}. Governed by this mechanism, the cutting-edge theoretical and experimental achievement of high-efficiency absorption bandwidths for MMAs are less than 30, to the best of our knowledge \cite{HuangSB2020SciBull, ZhouZL2022NSR, YangM2017MH, YeD2013PRL, ShangY2013IEEETAP, LinH2019NPhotonics, ZhouL2016SA, QuS2021PNAS, RileyCT2017PNAS, WangBX2023AFM, HuangY2018Nanoscale, CuiY2012NanoL}. Here, the ratio between the upper and lower bound of the operating band is used as a scale to evaluate the bandwidth. Additionally, the impact of complex higher-order propagating modes becomes significant in high-frequency regimes, bringing barriers to theoretical design and practical implementation. 
So far, achieving ultra-broadband high-efficiency MMAs over a hundred-fold bandwidth remains elusive.

In this study, we theoretically and experimentally demonstrate a seven-octave (128-fold-bandwidth) MMA. We introduce the quality-factor-weighted ($Q$-weighted) mode density as a fundamental concept to facilitate the achievement of ultra-broadband MMAs. The $Q$-weighted mode density incorporates the mode density, the resonant frequencies, radiative loss, and intrinsic loss of multiple resonant modes (MRM), manifesting a more fundamental and explicit connection with broadband absorption properties than mode density alone. It provides an alternative new pathway to develop ultra-broadband high-efficiency MMAs by modulating the $Q$-factors of a relatively small number of resonant modes, lifting the requirement of a huge number of resonant modes that are considerably challenging for MMAs in various wave systems. To demonstrate this, we introduce an acoustic system consisting of an array of cascade-parallel Helmholtz resonators and a covering wire mesh with micrometer-scale thickness. The resonator array provides a moderate mode density and the wire mesh enhances the $Q$-weighted mode density. Furthermore, we establish a theoretical framework and an experimental method to enable the design and measurement of ultra-broadband MMAs involving the consideration of high-order radiating waves. Finally, the presented MMA achieves high-efficiency absorption over 100-12800 Hz with an average absorption coefficient of 0.944 and a thickness of 35.3 cm approaching the causality-governed minimal thickness \cite{YangM2017MH,RozanovKN2000IEEETAP,QuS2021PNAS}. 
~\\

\section{Modulation based on Q-weighted mode density}

\begin{figure*}[htbp]
\centering
\includegraphics[width=12cm]{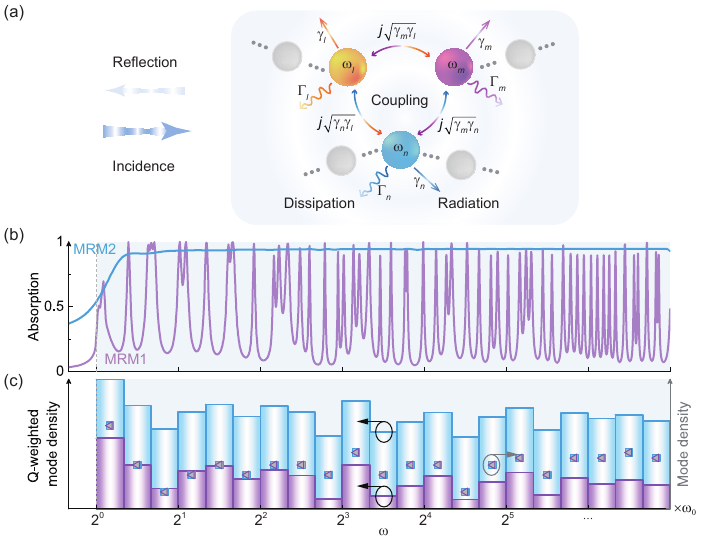}
\caption{\label{Fig.1} The physical mechanism of ultra-broadband MMAs based on MRM systems. (a), Schematic of the mode interactions of MRM systems. The three spheres with different colors represent three different modes. The gray spheres indicate the other different modes. ${\omega}$, ${\gamma}$, and ${\Gamma}$ represent resonant angular frequencies, the radiative loss, and intrinsic loss, respectively. The subscribes ($m$, $n$, and $l$) denote different modes.    $j\sqrt {{\gamma _l}{\gamma _m}}$ indicates the coupling between mode $l$ and mode $n$, as illustrated by the bi-directional arrows. The indications of other mode couplings follow a similar representation. (b), Absorption spectra of MRM1 (blue line) and MRM2 (purple line). (c), $Q$-weighted mode density (bar graphs, corresponding to the left axis) and mode density (blue squares and purple triangles scatter plots, corresponding to the right axis) of MRM1 (blue) and MRM2 (purple).
}
\end{figure*}

MRM systems offer exceptional potential for broadband wave control, thereby facilitating the achievement of ultra-broadband absorption \cite{FRH2020AM}. Through tuning the essential qualities of each resonant mode (resonant frequency, radiative loss, and intrinsic loss) and the coupling among the modes (Fig.~\ref{Fig.1}(a)), MRM systems can effectively suppress the dispersion of each resonant mode, thereby facilitating broadband absorption. Previous studies demonstrate that high mode density within the target band is required for broadband absorption \cite{YangM2017MH,ZhouZL2022NSR,FRH2020AM,GW2016OE}. The mode density is defined as 
${D_{\rm{m}}} = {N \mathord{\left/
 {\vphantom {N {{{\log }_2}\left( {{{{\omega _{\max }}} \mathord{\left/
 {\vphantom {{{\omega _{\max }}} {{\omega _{\min }}}}} \right.} {{\omega _{\min }}}}} \right)}}} \right.} {{{\log }_2}\left( {{{{\omega _{\max }}} \mathord{\left/
 {\vphantom {{{\omega _{\max }}} {{\omega _{\min }}}}} \right.} {{\omega _{\min }}}}} \right)}}$
, where $N$ is the number of resonant modes within the target frequency regime between ${\omega _{\max}}$  and ${\omega _{\min}}$. Accordingly, achieving large $D_{\rm{m}}$ within a broader frequency regime demands a greater number of resonant modes, while when it comes to an ultra-broadband frequency regime, it becomes rather challenging. Besides, due to the complex interactions among the modes within an ultra-wide bandwidth, only achieving high mode density cannot always ensure high-efficiency absorption. To overcome these challenges, we further take the other important qualities of resonant modes into consideration and present the concept of the $Q$-weighted mode density. The quality ($Q$) factor can be expressed as ${Q_n} = {{{\omega _n}} \mathord{\left/
 {\vphantom {{{\omega _n}} {\left( {2\left( {{\gamma _n} + {\Gamma _n}} \right)} \right)}}} \right.} {\left( {2\left( {{\gamma _n} + {\Gamma _n}} \right)} \right)}}$, where $n$ denotes the $n$-th mode, and $\omega _n$, $\gamma _n$ and $\Gamma _n$ represent the resonant angular frequency, radiative loss and intrinsic loss respectively. 
 In this study, the introduced $Q$-weighted mode density (${\chi _{{\rm{QMD}}}}$) is given by 
\begin{equation}
    {\chi _{{\rm{QMD}}}} = \frac{{\sum\limits_{n = 1}^N {{Q_n}^{ - 1}} }}{{{{\log }_2}\left( {{{{\omega _{\max }}} \mathord{\left/
 {\vphantom {{{\omega _{\max }}} {{\omega _{\min }}}}} \right.} {{\omega _{\min }}}}} \right)}},
\end{equation}
  which encompasses the mode density, the $Q$-factors, and the interactions of MRM. 
 
To demonstrate the design concept based on the $Q$-weighted mode density over the mode density, we employ coupled mode theory to calculate the absorption coefficient of two MRM systems with the same mode density within the same frequency band, where the absorption spectra are compared in Fig.~\ref{Fig.1}(b), Oscillations and numerous dips in the absorption spectrum can be observed for MRM1. In contrast, MRM2 exhibits significantly superior broadband absorption performance over MRM1 owing to the better modulation of the resonant modes' essential qualities and the improved interactions among the modes. Therefore, although the two systems possess identical $D_{\rm{m}}$ (indicated by purple triangles and blue squares in Fig.~\ref{Fig.1}(c)), the substantial differences in absorption spectra can arise from the different ${\chi _{{\rm{QMD}}}}$, where the MRM system with larger ${\chi _{{\rm{QMD}}}}$ (for MRM2) is capable of enhanced broadband absorption performance than that with smaller ${\chi _{{\rm{QMD}}}}$ (for MRM1), as depicted by the purple and blue bars in Fig.~\ref{Fig.1}(c). Consequently, by embracing the $Q$-factor, ${\chi _{{\rm{QMD}}}}$ fundamentally and comprehensively interprets the physical mechanisms and effect of mode couplings (see Fig. S1), leading to a more effective physical parameter for characterizing and designing the absorption properties of ultra-broadband MMAs based on MRM systems than $D_{\rm{m}}$. Specifically, to achieve the flat and high-efficiency absorption across multiple octaves of MRM2 (see the blue line in Fig.~\ref{Fig.1}(b)), there are two primary guidelines for the modulation of ${\chi _{{\rm{QMD}}}}$. The first guideline is that the ${\chi _{{\rm{QMD}}}}$ within the target frequency band require a high ${\chi _{{\rm{QMD}}}}$ to enable the MMAs to provide constantly high-efficient absorption. The second guideline is that MMAs should achieve a relatively higher ${\chi _{{\rm{QMD}}}}$ for the starting frequencies of the target frequency band to ensure a steep rise in the absorption curve, which is crucial for reducing the excessive responses and leading to a thinner structure \cite{ZhouZL2022NSR}. 
~\\

\section{Fundamental approaches for modulating broadband absorption performance}
\begin{figure*}[htbp]
\centering
\includegraphics[width=12cm]{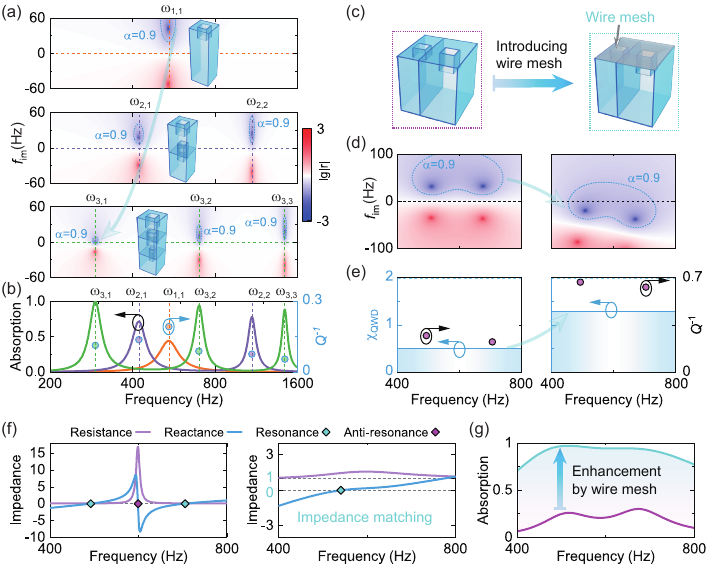}
\caption{\label{Fig.2} Approaches for enhancing ${\chi _{{\rm{QMD}}}}$ in acoustic system. (a), Acoustic frequency response of cascade structures. 
(b), Acoustic absorption spectra (solid lines, corresponding to the left axis) and ${Q^{ - 1}}$ at each resonance mode (dots, corresponding to the right axis) corresponding to each MCNEHR, where the curves and dots correspond to the real axes of the corresponding colors of the complex planes in (a). 
(c)-(g), Vertical division of the parallel structure and the change in the MMA performance by introducing wire mesh. (c), Illustration of the parallel structure (left) and the structure covered with wire mesh (right). (d), Complex frequency response of $\lg \left( {\left| r \right|} \right)$ corresponding to (c). (e), Distributions of ${Q^{ - 1}}$ (magenta dots, corresponding to the left axis) and ${\chi _{{\rm{QMD}}}}$ (blue bars, corresponding to the right axis) corresponding to (c). 
(f), Acoustic impedance corresponding to the structure before (left) and after (right) the introduction of the wire mesh. The rhombus indicates the location of resonance (cyan) and antiresonance (magenta). (g), Acoustic absorption spectra corresponding to the structure without (magenta) and with (cyan) the wire mesh.
} 
\end{figure*}

The above design concept is general for various wave systems. In the following, we demonstrate the concepts based on acoustic systems. Firstly, we need to construct acoustic resonators that can support an adequate number of resonant modes. Meanwhile, these resonators are anticipated to allow for high-degree-of-freedom modulations of the mode properties for achieving highly tunable mode couplings. 
In acoustics, Helmholtz resonators (HRs) are a classic implementation for introducing resonances \cite{HuangLJ2023NRP}. To obtain high-degree-of-freedom modulations of the mode properties, we propose the multilayer cascade neck-embedded Helmholtz resonator (MCNEHR) by transversely dividing the cavity into different sections using intermediate plates with embedded necks, based on the HR \cite{ZhouZL2022NSR}. The radiative and intrinsic loss of the resonant modes can be adjusted by the embedded necks (see Fig. S2(a)). The MCNEHRs facilitate the resonators to have resonant frequencies in the low-frequency regime and increase the number of modes (Fig.~\ref{Fig.2}(a) and Fig. S3), promoting high-efficiency absorption in low-frequency regime with sub-wavelength structures. 
However, for a single MCNEHR, the supported modes are typically uncoupled, and $Q^{-1}$ of the modes tends to decrease with the number of layers (dots in Fig.~\ref{Fig.2}(b)).
This characteristic of single MCNEHRs hinders effective manipulation of the ${\chi _{{\rm{QMD}}}}$, resulting in relatively narrow absorption bandwidths (see absorption spectra in Fig.~\ref{Fig.2}(b)). 

In addition, vertical spatial division (left panel of Fig.~\ref{Fig.2}(c)) can also increase the mode density, and more importantly, introduce the radiation coupling among different component resonators. The radiation coupling plays a crucial role in suppressing the dispersion nature of resonances (left panel of Fig.~\ref{Fig.2}(d)). However, it is difficult for resonators to have sufficient and controlled intrinsic loss across a wide bandwidth. When the intrinsic loss of the resonators are insufficient, $Q^{-1}$ and ${\chi _{{\rm{QMD}}}}$ will be inadequate (left panel of Fig.~\ref{Fig.2}(e)). In this case, the quasi-perfect absorption region (absorption efficiency exceeding 90 $\%$) will not fall on the real axis (left panel of Fig.~\ref{Fig.2}(d)), hindering the realization of broadband absorption. Furthermore, the presence of reflection zero points above the real axis indicates that the absorber includes redundant thickness \cite{YangM2017MH,ZhouZL2022NSR}.  Hence, although the transverse and vertical spatial divisions of HRs can effectively generate more resonant modes, sufficient intrinsic loss is required for optimal broadband absorption performance. Here, we present a simple approach to effectively modulate the intrinsic loss of the resonators across an extensive bandwidth by covering their surfaces with an extremely thin wire mesh (right panel of Fig.~\ref{Fig.2}(c), Fig. S2 and S4.). The additional intrinsic loss offered by the wire mesh can be adjusted by the mesh count ($C_{\rm{m}}$), facilitating the significantly enhanced $Q^{-1}$ and ${\chi _{{\rm{QMD}}}}$ (right panel of Fig.~\ref{Fig.2}(e)) and improved broadband performance (right panel of Fig.~\ref{Fig.2}(d), where the quasi-perfect absorption region intersects the real axis). Moreover, the reflection zeros are shifted to the lower half-plane, ensuring the absorber to achieve the minimum thickness determined by the causality constraint. 

The dispersive features of the resonators are evidently suppressed by adjusting the intrinsic loss supported by the wire mesh (Fig.~\ref{Fig.2}(f)). Without the aid of wire mesh, the resistance and reactance vary dramatically around the resonant frequency, exhibiting strong dispersion (left panel of Fig.~\ref{Fig.2}(f)). The introduction of wire mesh restricts the resistance and reactance within a low fluctuation range, enabling the MMA to accomplish broadband impedance matching, with the acoustic resistance nearly reaching 1 and the acoustic reactance approaching 0 (right panel of Fig.~\ref{Fig.2}(f)). Meanwhile, the suppression of anti-resonance is achieved. Consequently, the absorber's absorption performance is significantly improved in terms of both the bandwidth and the absorption coefficients (Fig.~\ref{Fig.2}(g)).
~\\

\section{Performance of seven-octave MMA}
\begin{figure*}[htbp]
\centering
\makebox[\textwidth][c]{\includegraphics[width=18cm]{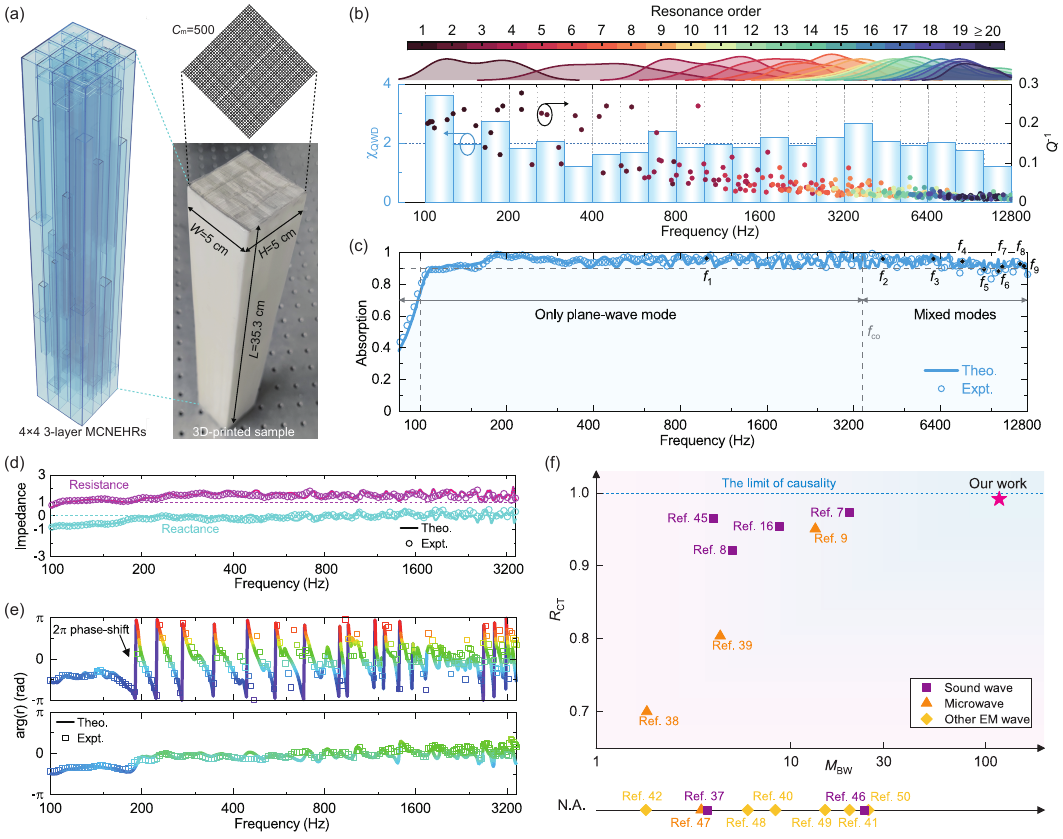}}
\caption{\label{Fig.3} Seven-octave metamaterial absorber. (a), Schematic of the seven-octave MMA, consisting of a 500-mesh wire mesh (upper right) covering the top surfaces of 16 parallel three-layer MCNEHR (left), with a 3D printed sample on the right panel.  (b), The ${\chi _{{\rm{QMD}}}}$ (bar charts) within each 1/3-octaves and the theoretical $Q^{-1}$ for resonant modes (colored dots). The distribution of different orders of resonant modes (color-filled curves). The absorption peak less than 0.1 is ignored. (c), Theoretical and experimental absorption spectra of the MMA. The labels $f_1$ to $f_9$ are points in frequency bands with different numbers of propagating modes. (d), Theoretical and experimental acoustic resistance and reactance spectra below ${f_{\rm{co}}}$. 
(e), Theoretical and experimental reflection phases of the plane-wave modes corresponding to the bare MMA (top) and the MMA covered with wire mesh (bottom). 
(f), Performance comparison of absorbers in different wave systems. Here, N.A. indicates that the $R_{\rm{CT}}$ is not available in the relevant work.} 
\end{figure*}

Guided by the aforementioned design strategies, we constructed an MMA employing only 16 three-layer MCNEHRs covered with a layer of wire mesh, with overall width $W = 5$ cm, height $H = 5$ cm (cutoff frequency ${f_{\rm{co}}} = 3432$ Hz), and thickness $L = 35.3$ cm (Fig.~\ref{Fig.3}(a)). 
Based on the developed theory (see Supplementary Note 1 and 2, Fig. S5 and S6), we performed global optimization to enhance the $Q$-weighted mode density and achieve extremely broadband high-efficiency acoustic absorption. 
The distribution of $Q^{-1}$ demonstrates that the MMA could generate multiple effective resonant modes over an extensive broadband range spanning seven octaves from 100 Hz to 12800 Hz (colored solid dots in Fig.~\ref{Fig.3}(b)).
After introducing the wire mesh, the ${\chi _{{\rm{QMD}}}}$ is significantly enhanced (blue bars in Fig.~\ref{Fig.3}(b) and Fig. S7(a)). Such a ${\chi _{{\rm{QMD}}}}$ distribution leads to extremely broadband high-efficiency absorption spanning seven octaves, exhibiting an average absorption coefficient of 0.944 (0.941 in the experiment) from 100 Hz to 12800 Hz (Fig.~\ref{Fig.3}(c)). Specifically, within 109-12800 Hz, the presented MMA realizes quasi-perfect absorption. The experimental and theoretical results are in good agreement, except for a slight difference in the high-frequency regime, which may be caused by the 3D printing imperfections. Furthermore, within the frequency regime below the waveguide's cut-off frequency ($f < {f_{{\rm{co}}}}$), the MMA exhibits perfect impedance matching and ideal over-damping characteristics (Fig.~\ref{Fig.3}(d)). Moreover, our presented MMA exhibited nearly perfect absorption performance in the frequency regime above the ${f_{{\rm{co}}}}$ (to be shown later). 
Under varying oblique incident angles, the MMA can also provide excellent absorption performance, indicating the robustness of the MMA's acoustic absorption performance and its insensitivity to incident angles (see Fig. S8).

From the results above, it can be observed that the wire mesh poses a significant impact on the overall absorption performance of the MMA despite its thin thickness of only 0.05 mm. Compared with the bare MMA (see Fig. S7 and S9), the addition of the wire mesh remarkably improves the intrinsic loss and the half-maximum relative bandwidth (i.e. $Q^{-1}$) of the MRM. This enhancement leads to increased ${\chi _{{\rm{QMD}}}}$, particularly in the low-frequency regime, effectively suppressing absorption dips caused by anti-resonances and impedance oscillations (Fig.~\ref{Fig.3}(d) and Fig. S7(b)). 
The effect of wire mesh can also be understood from the perspective of the phases of reflection coefficients. Without the wire mesh, numerous phase shifts of 0 to 2pi can be observed (top panel of Fig.~\ref{Fig.3}(e)). This phenomenon arises from the condition of the reflective zeros on the upper-half complex frequency plane (Fig. S10(a),(b)). 
In contrast, the MMA with wire mesh exhibits the minimum phase shift frequency dependence \cite{RozanovKN2000IEEETAP} throughout the band (bottom panel of Fig.~\ref{Fig.3}(e)).
 This result reveals that the zeros and poles reside on the same side of the real axis, indicating the absorber has reached the ${L_{\min }}$. Furthermore, we plotted the Riemann surface of the reflection coefficients of the plane-wave modes over the entire target band and the reflection phases on their real axes, which further verifies this point (Fig. S10).

To better evaluate the performance of the presented MMA, we introduce two quantitative indicators, the bandwidth multiple (${M_{{\rm{BW}}}}$) and the causality thickness ratio (${R_{{\rm{CT}}}}$), for comprehensively comparing the absorbers in various wave systems.
The ${M_{{\rm{BW}}}}$ is defined as ${M_{{\rm{BW}}}} = {f_{\max }}/{f_{\min }}$, where ${f_{\max}}$ and ${f_{\min}}$ respectively represent the maximum and minimum frequencies of the frequency band consistently exhibiting quai-perfect absorption. For the MMA presented in this work, the quasi-perfect absorption band ranges from 109 Hz to 12800 Hz, resulting in a ${M_{{\rm{BW}}}} = 117$. Besides, the ${R_{{\rm{CT}}}}$ is defined as ${R_{{\rm{CT}}}}{{ = {L_{{\rm{min}}}}} \mathord{\left/
 {\vphantom {{ = {L_{{\rm{min}}}}} L}} \right.} L}$, where $L$ represents the sample thickness. 
For our presented MMA, $L = 35.3$ cm and ${L_{\min }}=35$ cm, and thus ${R_{{\rm{CT}}}}{\rm{ = 0.991}}$. It is emphasized that the ${R_{{\rm{CT}}}}$ can only reach the maximum value 1 limited by the causal constraint when $L = {L_{\min }}$, i.e., the thickness of the absorber reaches the minimum thickness of the causal domination (blue dash line in Fig.~\ref{Fig.3}(f)). Therefore, ${R_{{\rm{CT}}}}$ precisely highlights the optimal trade-off between absorption properties and the thickness of MMAs. And ${M_{{\rm{BW}}}}$ carries the importance of the effective bandwidth. These two indicators enable a comprehensive evaluation of the absorber's overall performance. Compared to other absorbers of various wave systems (Fig.~\ref{Fig.3}(f) and Fig. S11) \cite{Qusc2022SA,YangM2017MH,ZhouZL2022NSR,DingH2022IJMS,HuangSB2020SciBull,QS2024MTP,QuS2021PNAS,ShangY2013IEEETAP,YeD2013PRL,ZJ2016JEM,LY2022NanoL,LinH2019NPhotonics,ChowdharyAK2021JOSAB,AQ2023AM,ZhouL2016SA,RileyCT2017PNAS}, our presented MMA boasts an order of magnitude higher ${M_{{\rm{BW}}}}$ and mostly reaches the physical limit of ${R_{{\rm{CT}}}}$ (see table S1 for detailed parameters).
~\\

\section{Absorption performance under mixed-mode incidence}
\begin{figure*}[htbp]
\centering
\makebox[\textwidth][c]{\includegraphics[width=18cm]{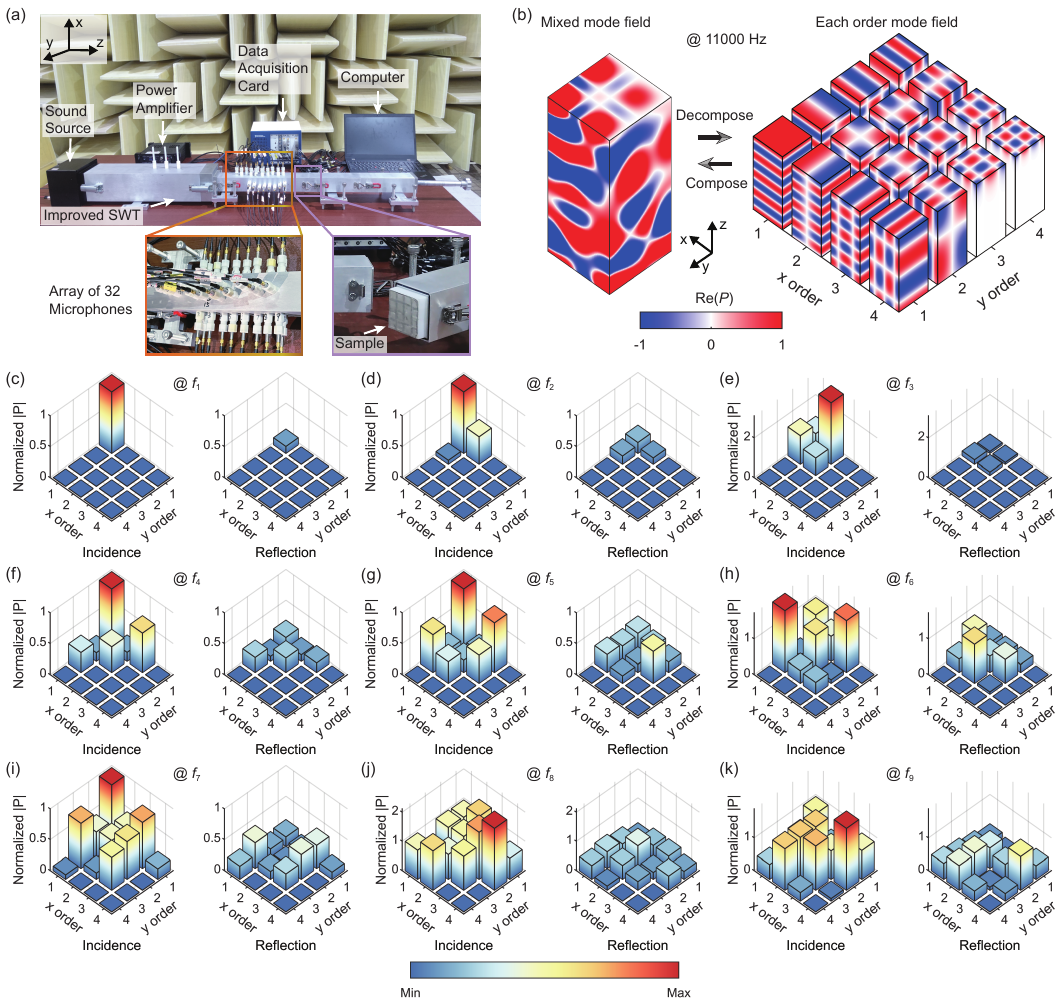}}
\caption{\label{Fig.4} Absorption performance of MMA for mixed modes.  (a), Schematic of the experimental platform in the anechoic chamber. 32 microphones are placed in the measurement section, which supports mode decomposition of the highest order of 16. (b), The compose and decompose incident field at 11000 Hz consisting of simultaneous plane wave modes (only 1), high-order propagation modes (12 in total), and evanescent wave modes (3 in total).  
(c)-(k), Incident and reflected mode amplitudes from experimental results corresponding to frequency points $f_1$ to $f_9$ (with 1, 3, 4, 6, 8, 9, 11, 13, and 15 propagation modes, respectively) in the frequency bands with different numbers of propagation modes in Fig.~\ref{Fig.3}(c). 
} 
\end{figure*}

Due to the inadequacies in previous theories and challenges in practical implementation, most research on acoustic MMAs has steered clear of investigating mixed-mode absorption properties, significantly hindering the advancement of ultra-broadband acoustic MMAs. To overcome this bottleneck, a more robust theoretical design framework and a more comprehensive experimental testing platform (Fig.~\ref{Fig.4}(a), the multi-mode standing wave tube (SWT)) have been developed.
The platform employs 32 microphones to decompose the mixed modes above the ${f_{\rm{co}}}$, enabling the calculation of the absorption coefficients for higher-order propagating modes (see Fig. S12). In addition, our developed theory enables accurate calculation of the absorption performances of MMAs beyond the ${f_{\rm{co}}}$ limitation. As a result, the experimental results fit well with the theoretical predictions (Fig.~\ref{Fig.3}(c)).  

The propagating modes at different frequencies are determined by the dispersion relation (Supplementary Note 6). For instance, at 11,000 Hz (Fig.~\ref{Fig.4}(b)), the plane wave modes of order (1,1) and 12 higher-order modes ranging from orders (1,2) to (3,3) exist.
The remaining higher-order modes are all evanescent modes, as illustrated by orders (3,4), (4,3), and (4,4) in Fig. 4(b)). Based on this classification, the entire operational bandwidth can be divided into nine sub-bands with varying numbers of propagating modes (see Fig. S13). The overall absorption performance of the mixed modes is demonstrated in Fig.~\ref{Fig.3}(c) (above the ${f_{\rm{co}}}$). Further, to provide a clearer illustration of the absorption of the mixed modes, a frequency point was selected from each of the nine frequency bands (${f_{1}}$ to ${f_{9}}$) and decomposed using the aforementioned measurements. 
Figures ~\ref{Fig.4}(c)-(k) are provided, displaying the measured incident (left panel) and reflected (right panel) mode amplitudes for each order of the propagating modes from ${f_{1}}$ to ${f_{9}}$ (see Fig. S14 for mode energy flow). These figures clearly demonstrate the highly robust absorption performance of our presented MMA under the incidences of different mixed modes. Nearly perfect absorption is observed for the majority of modes at various frequencies, with only trivial energy transformation occurring between different modes (see Fig. S15 for the mode transformation diagrams). This energy transformation results in a small number of higher-order modes exhibiting a reflection amplitude slightly larger than the incidence amplitude, such as (1,2) mode in ${f_{2}}$ (Fig.~\ref{Fig.4}(d)) and (3,2) mode in ${f_{6}}$ (Fig.~\ref{Fig.4}(h)). However, these modes only pose negligible influence on the overall absorption performance, because the plane-wave modes ((1,1) in Fig.~\ref{Fig.4}(c)-(k)) with the highest energy share are significantly absorbed.
~\\

\section{conslusions}

In conclusion, our metamaterial absorber (MMA) achieves ultra-broadband absorption across an unprecedented seven-octave range with remarkable efficiency. By leveraging quality-factor-weighted mode density to optimize resonant modes and manage intrinsic losses, we enable effective absorption from 100 Hz to 12,800 Hz. This innovative approach not only advances the performance of metamaterial absorbers but also establishes a robust framework for extending ultra-broadband applications to diverse wave systems, including non-Hermitian systems and novel topological devices. Moreover, the methodology shows significant potential for new applications, such as in aerospace engines, where ultra-broadband absorption could greatly enhance performance and efficiency.

~\\



\section*{Acknowledgments}

This work is supported by the National Natural Science Foundation of China under Grant Nos. 12074286, 92263208 and 62375232,  the Shanghai Science and Technology Committee under Grant No. 21JC1405600, the University Grants Committee/Research Grants Council of the Hong Kong Special Administrative Region, China under Grant Nos. AoE/P-502/20, C5031-22G, CityU15303521, CityU11305223 and G-CityU 101/22, and City University of Hong Kong under Grant Nos. 9380131 and 7005867. 
~\\

\nolinenumbers 
%

\end{document}